\input harvmac
\input epsf
\Title{\vbox{\hbox{HUTP-98/A017, NUB 3170}}}
{\vbox{\hbox{\centerline{String Expansion as Large $N$ Expansion of Gauge
Theories}}
\vskip .3in
\hbox {\centerline{ }}}}
\vskip .3in
\centerline{Michael Bershadsky$^1$,
Zurab Kakushadze$^{1,2}$ and
Cumrun Vafa$^1$}
\vskip .2in
\centerline{\it $^1$Lyman Laboratory of Physics, Harvard
University, Cambridge, MA 02138}
\centerline{\it $^2$Department of Physics, Northeastern University, Boston, MA
02115}

\vskip .3in
\centerline{\bf Abstract}
We consider
string perturbative expansion in the
presence of D-branes imbedded in orbifolded space-time.  In
the regime where the string coupling is weak and $\alpha'\rightarrow 0$,
the string perturbative expansion coincides with `t Hooft's
large $N$ expansion.  We specifically concentrate on theories
with $d=4$ and ${\cal N}=0,1,2,4$, and
use world-sheet orbifold techniques to prove
vanishing theorems for the field theory $\beta$-functions to all
orders in perturbation theory in the large $N$ limit.
This is in accord with recent predictions.

\vskip .2in
\Date{March 8, 1998}

\newsec{Introduction}
One of the most interesting ideas in gaining insight into
the structure of gauge theory is `t Hooft's idea of considering
a large $N$ limit \ref\tho{G. `t Hooft, ``A Planar Diagram Theory For Strong
Interactions'', Nucl. Phys. {\bf 72}
(1974) 461.}.  It was noticed in \tho\
that in this limit the gauge theory diagrams organize themselves
in terms of Riemann surfaces, where addition
of each extra handle on the surface corresponds to suppression
by $1\over N^2$.  In fact, this similarity led `t Hooft to speculate
about possible connections with the ``dual'' model perturbation
expansion--which today we call string theory perturbation.

Even though the idea sounded promising, no direct connection
between the two was made for a long time.
The first concrete connection came with the beautiful work of
Witten
\ref\css{E. Witten, ``Chern-Simons Gauge Theory As A String Theory'',
hep-th/9207094.}\ where it was shown that at least in the
context of topological strings, with boundaries mapped to topological
versions of $N$ D3-branes,
the string expansion was
actually the same as the large $N$ expansion of the 3d
Chern-Simons gauge
theory.  In particular, diagram by diagram string theory expansion
was mapped to large $N$ expansion of gauge theory, in the sense
of `t Hooft.

In a seemingly unrelated development, it was noticed
\ref\kos{D.A. Kosower, B.-H. Lee and V.P. Nair, ``Multi Gluon Scattering:
A String Based Calculation'', Phys. Lett. {\bf B201} (1988) 85\semi
Z. Bern and D.A. Kosower, ``Efficient Calculation of One Loop QCD Amplitudes'',
Phys. Rev. Lett. {\bf 66} (1991) 1669.} (for a
recent discussion, see, {\it e.g.}, \ref\mreof{
Z. Bern, L. Dixon, D.C. Dunbar, M. Perelstin and J.S. Rozowsky, ``On the
Relationship between Yang-Mills Theory and Gravity and Its Implication for
Ultraviolet Divergences'', hep-th/9802162.}) that
string theory perturbation techniques is a very useful way of
summing up various field theory Feynman diagrams.  The basic
idea here is to consider a limit $\alpha'\rightarrow 0$,
where string theory reduces to its massless modes, and try to extract
the contribution of gauge fields and matter in the corresponding
string theory diagram.  This direction has been studied extensively,
with various interesting applications.  These applications suggest
that even if we are just interested in gauge theories, the string
perturbation techniques are very powerful and should not be overlooked.

The basic idea of this paper is to combine these two approaches.
Namely, we consider type II strings in the presence of a large number
$N$ of D-branes and consider a limit where $\alpha'\rightarrow 0$
while keeping $\lambda =N\lambda_s$ fixed, where $\lambda_s$ is the type II
string coupling.  Note that in this context a world-sheet with $g$
handles and $b$ boundaries is weighted with
$$(N\lambda_s)^b \lambda_s^{2g-2}=\lambda^{2g-2+b}N^{-2g+2}~.$$
After we identify
$\lambda_s =g_{YM}^2$, this is the same as large $N$ expansion considered
by `t Hooft.  Note that for this expansion to make sense we have
to consider the limit where $N\rightarrow \infty$ while fixing
$\lambda$ at a small value $\lambda <1$.  If $\lambda >1$  then
no matter how large $N$ is, for sufficiently many boundaries the
higher genus terms would be relevant, and we lose the genus expansion
of large $N$.  This, of course, is the same as the domain of validity
of string perturbation theory.  Therefore, one should expect
a surface expansion in large $N$ {\it only} for small $\lambda$.
In this limit we can map the string diagrams directly to
(specific sums of) large $N$ Feynman diagrams.
Note in particular that the genus $g=0$ planar diagrams
dominate in the large $N$ limit.

In the large $N$ limit we are still left with the free parameter
$\lambda$ and there are
two natural regimes of parameters
to consider.  As just discussed the case which makes
contact with large $N$ analysis of `t Hooft and string perturbation theory is
small $\lambda$.  It is also interesting to ask what happens for large
$\lambda$.  This is a very non-trivial question and is beyond the domain
of validity of `t Hooft's large $N$ analysis or string theory perturbation
techniques.
This is precisely the case considered recently in several very
interesting papers
\ref\kleb{I.R. Klebanov, ``Worldvolume Approach to Absorption by
Nondilatonic Branes'',
Nucl. Phys. {\bf B496} (1997) 231, hep-th/9702076.}\ref\gub{S.S. Gubser
and I.R. Klebanov, ``Absorption by Branes and Schwinger Terms in
the Worldvolume Theory'', Phys. Lett. {\bf B413} (1997) 41,
hep-th/9708005.}\ref\malda{J.M. Maldacena, ``The Large $N$ Limit of
Superconformal Field
Theories and Supergravity'', hep-th/9711200.}\ref\poly{S.S.  Gubser,
I.R. Klebanov and A.M.
Polyakov, ``Gauge Theory Correlators from
Non-Critical String Theory``, hep-th/9802109.}\ref\Oog{G.T. Horowitz and
H. Ooguri, ``Spectrum of Large $N$ Gauge
Theory from Supergravity'', hep-th/9802116.}\ref\witt{E. Witten, ``Anti-de
Sitter
Space And Holography'', hep-th/9802150.},
and has also been studied in related
works \ref\addi{S.S. Gubser, I.R. Klebanov and A.W. Peet, ``Entropy and
Temperature of Black 3-branes'', Phys. Rev {\bf D54} (1996) 3915,
hep-th/9602135\semi
S.S. Gubser, I.R. Klebanov and A.A. Tseytlin, ``String Theory and Classical
Absorption
by Three Branes'', Nucl. Phys. {\bf B499} (1997) 217, hep-th/9703040\semi
J.M. Maldacena and A. Strominger, ``Semiclassical Decay of Near Extremal
Fivebranes'', hep-th/9710014\semi
A.M. Polyakov, ``String Theory and Quark
Confinement'', hep-th/9711002\semi N. Itzhaki, J.M. Maldacena,
J. Sonnenschein and S. Yankielowicz,
``Supergravity and the Large $N$ Limit of Theories With Sixteen Supercharges'',
hep-th/9802042\semi
S. Ferrara and C. Fronsdal, ``Conformal Maxwell Theory as a Singleton
Field Theory on $ADS_5$, IIB Three Branes and Duality'', hep-th/9712239\semi
M. Berkooz, ``A Supergravity Dual of a (1,0) Field Theory in Six Dimensions'',
hep-th/9802195\semi V. Balasubramanian and F. Larsen, ``Near Horizon Geometry
and Black Holes
in Four Dimensions'', hep-th/9802198\semi S.-J. Rey and J. Yee, ``Macroscopic
Strings as Heavy Quarks of Large $N$ Gauge
Theory and Anti-de Sitter Supergravity'', hep-th/9803001\semi
J.M. Maldacena, ``Wilson loops in large $N$ field theories'',
hep-th/9803002\semi
S.S.  Gubser, A. Hashimoto, I.R. Klebanov and M. Krasnitz, ``Scalar Absorption
and the Breaking of the Worldvolume Conformal Invariance'', hep-th/9803023\semi
I.Ya. Aref'eva and I.V. Volovich, ``On Large $N$ Conformal Theories, Field
Theories in
the Anti-de Sitter space and Singletons'', hep-th/9803023\semi
L. Castellani, A. Ceresole, R. D'Auria, S. Ferrara, P. Fr{\'e} and M.
Trigiante,
``$G/H$ M-branes and $AdS_{p+2}$ Geometries'', hep-th/9803039\semi
O. Aharony, Y. Oz and Z. Yin, ``M Theory on $AdS_p\times S^{11-p}$ and
Superconformal Field Theories'', hep-th/9803051\semi
S. Ferrara, C. Fronsdal and A. Zaffaroni,
``On N=8 Supergravity on $AdS_5$ and N=4 Superconformal Yang-Mills theory'',
hep-th/9802203.}.
This is a limit where one expects
an effective supergravity description to take over.

We will be considering the limit where $\lambda$ is small.
We shall see that the string world-sheet techniques
allow us to prove certain statements about the gauge theory in this limit,
which in principle should be properties of Feynman diagrams.  But since
string theory organizes Feynman diagrams in a very economical way, it turns
out that the proof is obvious only in the string theory setup.   Using string
theory perturbation techniques we establish part of the
conjectures in \ref\lnv{A. Lawrence, N. Nekrasov and C. Vafa,
``On Conformal Theories in Four Dimensions'', hep-th/9803015.}\ (extending
the work in \ref\ks{S. Kachru and E. Silverstein, ``4d Conformal Field Theories
and Strings on Orbifolds'', hep-th/9802183.}) in connection
with constructing four dimensional conformal field theories\foot{For a review
of field theory discussions of this subject, see, {\it e.g.}, \ref\stras{M.J.
Strassler,
``Manifolds of Fixed Points and Duality in Supersymmetric Gauge Theories'',
Prog. Theor. Phys. Suppl. {\bf 123} (1996) 373, hep-th/9602021.}. For a
recent attempt for constructing finite gauge theories via orientifolds, see,
{\it e.g.},
\ref\iba{L.E. Ib{\'a}{\~n}ez, ``A Chiral D=4, N=1 String Vacuum with a Finite
Low
Energy Effective Field Theory'', hep-th/9802103.}.}
(including the case with no supersymmetries).  More precisely,
we {\it prove} that in the large $N$ limit the $\beta$-functions of
all theories considered in \lnv\ are identically zero.  We
also gain insight into possible $1/N$ corrections in this context.  The
vanishing of the $\beta$-functions
was proved up to two loops in the ${\cal N}
=1$ examples in \lnv, and up to one-loop level in the ${\cal N}=0$
case.  Given how cumbersome such Feynman diagram computations
are it is quite pleasant to observe the power of string perturbation
techniques (in the context of orbifolds) in proving such statements.
Moreover, in doing so we also gain insight into the
conditions imposed in the orbifold construction in \lnv\
(and, in particular, {\it why} the orbifold groups considered
in \lnv\ should act in
the regular representation when acting on
gauge degrees of freedom).  In fact we will be able to show more.
Namely, we show that any correlation computation for these theories
reduces as $N\rightarrow \infty$ to the corresponding computation
in the ${\cal N}=4$ theory.  We will also see why the ${1\over N}$
corrections will be different from those of the ${\cal N}=4 $ theory.

The remainder of this paper is organized as follows. In section 2 we
describe construction of gauge theories via type IIB D3-branes in
orbifold backgrounds
which lead to theories that are (super)conformal in the large $N$ limit.
In section 3 we show that in these theories the
$\beta$-functions as well as anomalous scaling dimensions vanish
to all orders in the large $N$ limit. In section 4 we discuss subleading
corrections at large $N$. We point out that, subject to certain assumptions,
one may also be able to prove that
${\cal N}=1$ theories may be superconformal even at finite $N$.
We also discuss the issues that need to be understood for checking
such statements for the ${\cal N}=0$ case.

\newsec{Setup}

We start with type IIB string theory with $N$ parallel D3-branes where
the space transverse to the
D-branes is ${\cal M}={\bf R}^6/\Gamma$.
The orbifold group
$\Gamma= \left\{ g_a \mid a=1,\dots,
|\Gamma| \right\}$ ($g_1=1$)
must be a finite discrete subgroup of $Spin(6)$.
If $\Gamma\subset SU(3)$ ($SU(2)$), we have
${\cal N}=1$ (${\cal N}=2$) unbroken supersymmetry,
and ${\cal N}=0$, otherwise.

We will confine our attention to the cases where type IIB on ${\cal M}$ is
a modular invariant theory\foot{This is always the case in the
supersymmetric case.  For the non-supersymmetric case this is also
true if
$\not\exists{\bf Z}_2\subset\Gamma$. If $\exists{\bf Z}_2\subset\Gamma$,
then modular invariance requires that the set of points in ${\bf R}^6$
fixed under the ${\bf Z}_2$ twist has real dimension 2.}. The action of the
orbifold on
the coordinates $X_i$ ($i=1,\dots,6$) on ${\cal M}$ can be described
in terms of $SO(6)$ matrices:
$g_a:X_i\rightarrow (g_a)_{ij} X_j$.
The world-sheet fermionic superpartners of $X_i$
transform in the same way.
We also need to specify
the action of the orbifold group on the Chan-Paton charges carried by the
D3-branes. It is described by $N\times N$ matrices $\gamma_a$ that
form a representation of $\Gamma$. Note that $\gamma_1$ is the identity
matrix and ${\rm Tr}(\gamma_1)=N$.

The D-brane sector of the theory is described by an {\it oriented} open
string theory.   In particular the world-sheet expansion corresponds
to summing over oriented Riemann surfaces with arbitrary genus $g$ and
arbitrary
number of boundaries $b$, where the boundaries of the world-sheet are mapped
to the D3-brane world-volume.  Moreover we consider various ``twists''
corresponding to orbifold sectors, around the cycles of the Riemann
surface.  The choice of ``twists''
corresponds to a choice of homomorphism of the fundamental group of
the Riemann surface with boundaries to $\Gamma$.

For example, consider one-loop vacuum amplitude ($g=0,b=2$). The
corresponding graph is an annulus whose boundaries lie on D3-branes.
The one-loop partition function in the
light-cone gauge is given by
\eqn\partition{
 {\cal Z}={1\over 2|\Gamma|}\sum_a
 {\rm Tr}  \left( g_a (1+(-1)^F)
 e^{-2\pi tL_0}
 \right)~,
}
where $F$ is the fermion number operator, $t$ is the real modular parameter
on the cylinder, and the trace includes sum over the Chan-Paton factors.
The states in the Neveu-Schwarz (NS) sector are space-time bosons and enter the
partition function with weight $+1$, whereas the states in the Ramond (R)
sector are
space-time fermions and contribute with weight $-1$.

Note that the elements $g_a$ acting in the Hilbert space of open strings
will act both on the left-end and the right-end of the open string.
In particular $g_a$ corresponds to $\gamma_a\otimes \gamma_a$ acting
on the Chan-Paton indices.
The individual terms in the sum in \partition\ therefore
have the following form:
\eqn\gsq{
 \left({\rm Tr}(\gamma_a)\right)^2 {\cal Z}_a~,
}
where ${\cal Z}_a$ are characters
corresponding to the world-sheet degrees of freedom. The ``untwisted''
character
${\cal Z}_1$ is the same as in the ${\cal N}=4$ theory for which
$\Gamma=\{1\}$. The information about the fact that the orbifold theory
has reduced supersymmetry is encoded in the ``twisted'' characters
${\cal Z}_a$, $a\not=1$.

Here we are interested in constructing finite gauge theories. Since
${\cal N}=4$ gauge theories are finite, we can hope to obtain
finite gauge theories (at least in the large $N$ limit)
with reduced supersymmetry by arranging for
the twisted sector contributions to the $\beta$-functions and anomalous
scaling dimensions to be absent. The above discussion suggests one natural
way of possibly achieving this goal. Consider representations of $\Gamma$
such that
\eqn\trace{
 {\rm Tr}(\gamma_a)=0~\forall a\not=1~.
}
In section 3 we will show that gauge theories corresponding to such
representations are indeed finite in the large $N$ limit to all orders in
perturbation theory.   In fact, we will see in the next subsection
that cancellation of tadpoles in the closed string channel requires this
trace condition.  Moreover, we will show that this trace condition
fixes the representation $\gamma$ to be a sum of copies of the regular
representation of $\Gamma$.

\subsec{Tadpole cancellation}

In this section we investigate the conditions under which the oriented open
string
theory that describes the D-brane sector in the above framework is finite in
the
{\it ultraviolet}. Ultraviolet finiteness guarantees absence of anomalies, and
is related to tadpole cancellation in the closed
string channel. Here we discuss the one-loop tadpole cancellation
conditions which as we will see impose constraints on the Chan-Paton matrices
$\gamma_a$.

The characters ${\cal Z}_a$ in \gsq\ are given by
\eqn\terms{
 {\cal Z}_a =
 \left[\eta(e^{-2\pi t})\right]^{-2-d_a}
 \left({\cal X}_a (e^{-2\pi t})-{\cal Y}_a (e^{-2\pi t}) \right)~.
}
Here $d_a$ is the real dimension of the set of points fixed
under the twist $g_a$. Two of the $\eta$-functions come from
the oscillators corresponding to the space-time directions
filled by D3-branes (and the time-like and longitudinal contributions are
absent due to the light-cone gauge). The other $d_a$
$\eta$-functions come from the oscillators corresponding
to the directions transverse to the D-branes. Finally, the
characters ${\cal X}_a$, ${\cal Y}_a$ correspond to the contributions of
the world-sheet fermions, as well as the
world-sheet bosons with $g_a$ acting non-trivially
on them (for $a\not=1$):
${\cal X}_a$ arises in the trace ${\rm Tr}\left(g_a e^{-2\pi tL_0}\right)$,
whereas
${\cal Y}_a$ arises in the trace ${\rm Tr}\left(g_a (-1)^F e^{-2\pi
tL_0}\right)$
(see \partition). We will not need their explicit form here.

The contributions to the one-loop vacuum amplitude corresponding to
${\cal Z}_a$ are (proportional to)
$$\int _0^\infty {dt\over t^3} ~ ~\left({\rm Tr}(\gamma_a)\right)^2 {\cal Z}_a
=
 {\cal A}_a - {\cal B}_b ~,$$
where
\eqn\oneloop{\eqalign{
 & {\cal A}_a =\left({\rm Tr}(\gamma_a)\right)^2 \int_0^\infty {dt\over t^3} ~
 \left[\eta(e^{-2\pi t})\right]^{-2-d_a}
 {\cal X}_a (e^{-2\pi t})~,\cr
 & {\cal B}_a =\left({\rm Tr}(\gamma_a)\right)^2 \int_0^\infty {dt\over t^3} ~
 \left[\eta(e^{-2\pi t})\right]^{-2-d_a}
 {\cal Y}_a (e^{-2\pi t})~.
}}
These integrals\foot{For space-time supersymmetric
theories the total tadpoles vanish: ${\cal A}_a-{\cal B}_a=0$. (The entire
partition function vanishes as the numbers of
space-time bosons and fermions are equal.) For consistency, however, we must
extract individual contributions ${\cal A}_a$ and ${\cal B}_a$
and make sure that they cancel as well. Thus, cancellation of the tadpoles in
${\cal B}_a$
is required for consistency of the untwisted and twisted R-R four-form (to
which
D3-branes couple) equations of motion.}
are generically divergent as $t\rightarrow 0$ reflecting
presence of tadpoles.
To extract these divergences we can change variables $t=1/\ell$ so that the
divergences correspond to $\ell\rightarrow\infty$:
\eqn\newone{\eqalign{
 & {\cal A}_a =\left({\rm Tr}(\gamma_a)\right)^2 \int_0^\infty
 {d\ell\over \ell^{d_a/2}}
  \sum_{\sigma_a} A_{\sigma_a} e^{-2\pi\ell \sigma_a}~,\cr
 & {\cal B}_a =\left({\rm Tr}(\gamma_a)\right)^2 \int_0^\infty
  {d\ell\over \ell^{d_a/2}}
 \sum_{\rho_a} B_{\rho_a}e^{-2\pi\ell \rho_a}~.
}}
The closed string states contributing to ${\cal A}_a$ (${\cal B}_a$)
in the transverse channel
are the NS-NS (R-R) states with $L_0={\overline L}_0=\sigma_a (\rho_a)$
(and $A_{\sigma_a} (B_{\rho_a})>0$ is
the number of such states). Note that the massive states with
$\sigma_a(\rho_a)>0$
do not lead to divergences as $\ell\rightarrow\infty$.
The ground states in the
R-R sectors are massless.
Note that in $\ell\rightarrow\infty$ limit
the value of $d_a$ in the
prefactor $\ell^{-d_a/2}$ determines the divergence property
of the integral.  Note that $d_a$ is the real dimension of the set of points
fixed
under the twist $g_a$.
We thus get divergences in ${\cal B}_a$ for large $\ell$
for $d_a\leq 2$.  In such a case we must make sure that the corresponding
${\rm Tr}(\gamma_a)=0$.
Given the orientability of $\Gamma$ the allowed values of $d_a$
are $0,2,4,6$.
 For $d_a=0,2$ tadpole cancellations thus
require ${\rm Tr}(\gamma_a)=0$.
For the case $d_a=4$ the
corresponding twisted NS-NS closed string sector will contain tachyons. This
will lead to
a tachyonic tadpole in ${\cal A}_a$ unless ${\rm Tr}(\gamma_a)=0$ in the $g_a$
twisted
sector.  For $d_1=6$, there is no divergence in ${\cal B}_1$ and
thus we have no restriction
for ${\rm Tr}(\gamma_1)=N$.   We therefore conclude that to cancel all tadpoles
it is necessary that
\eqn\pole{
 {\rm Tr}(\gamma_a )=0~\forall a\not=1~.
}
Since the untwisted NS-NS closed string sector does not contain tachyons, no
further constraint arises on ${\rm Tr}(\gamma_1 )=N$ (so that the number of
D3-branes
is arbitrary). Thus we conclude that one-loop tadpole cancellation is possible
if and
only if the constraint \pole\ is satisfied. This is precisely the condition
\trace\
discussed in the beginning of this section.

Thus, we see that the condition on ${\rm Tr}(\gamma_a )$ is necessary
and sufficient to guarantee one-loop
ultraviolet finiteness and consistency of the corresponding string theory.
For illustrative purposes, to see what can go wrong if we relax this condition,
let us consider the following example. Let ${\cal M}={\bf C}^3/{\Gamma}$, where
the action of $\Gamma$ on the complex coordinates $X_\alpha$ ($\alpha=1,2,3$)
on ${\cal M}$ is that of the Z-orbifold: $g:X_\alpha\rightarrow \omega
X_\alpha$
(where $g$ is the generator of $\Gamma$, and $\omega=e^{2\pi i/3}$). Next, let
us
choose the representation of $\Gamma$ when acting on the Chan-Paton charges
as follows: $\gamma_g={\rm diag}({\bf I}_{N_1},\omega {\bf I}_{N_2},
\omega^2 {\bf I}_{N_3})$
(where $N_1+N_2+N_3=N$, and ${\bf I}_m$ is an $m\times m$ identity matrix).
The massless spectrum of this model is ${\cal N}=1$
$U(N_1)\otimes U(N_2)\otimes U(N_3)$ gauge theory with the following matter
content:
$$ 3({\bf N}_1,{\overline {\bf N}}_2,{\bf 1})~,~
     3({\bf 1}, {\bf N}_2,{\overline {\bf N}}_3)~,~
     3({\overline {\bf N}}_1,{\bf 1},{\bf N}_3)~.$$
Note that this spectrum is anomalous (the non-Abelian gauge anomaly does not
cancel)
unless $N_1=N_2=N_3$. On the other hand, ${\rm Tr}(\gamma_g )=0$ if and only if
$N_1=N_2=N_3$. Thus, we see that the constraint \pole\ derived from the
tadpole
cancellation conditions is necessary in this case to have a consistent gauge
theory.

Here we should mention that not all the choices of $\gamma_a$ that do not
satisfy
\pole\ lead to such apparent inconsistencies. Consider for example the
same $\Gamma$ as above but with $\gamma_g={\bf I}_N$. The massless spectrum of
this
model is ${\cal N}=1$ $U(N)$ gauge theory with no matter, so it is anomaly
free.  It would be interesting to see to what extent such theories can
be studied using the present string theory construction.

\subsec{Regular representation}

In the examples constructed in \lnv,  generalizing
the construction of \ks\ in attempts for constructing conformal field theories
in
four dimensions, the orbifold action on
the gauge degrees of freedom
was chosen to be an $n$-fold copy of the {\it regular}
representation of $\Gamma$. Here we prove that
tadpole cancellation conditions
for $\gamma_a$ \pole\
 correspond to having an $n$-fold copy of the regular representation of
$\Gamma$.
Conversely, if $\gamma_a$ form an $n$-fold regular representation of $\Gamma$,
then the condition \pole\ is satisfied.

{}Recall that the regular representation corresponds
to considering the vector space which is identified with
$\{ |g_a\rangle \} $ for elements $g_a\in \Gamma$.
The action of the group in the regular representation is given by
$$\gamma_b |g_a\rangle =|g_a g_b\rangle~.$$
Let us consider $g_b\not=1$.  Then it is clear that in this
representation we have ${\rm Tr}(\gamma_b)=0$ since for all
$g_a$ we have $g_a g_b\not= g_a$. Also note that
${\rm Tr}(\gamma_1)=|\Gamma|$.
If we consider $n$ copies of this representation we will have the condition
that the trace of all elements are zero except for the identity element
whose trace is $n|\Gamma|$. This is the same condition as \trace.

Next, we show that \pole\ implies that $\gamma_a$ form an $n$-fold
copy of regular representation of $\Gamma$.
First, let us establish
that ${\rm Tr}(\gamma_1)=n |\Gamma| $ for some integer $n$,
{\it i.e.}, that the dimension of the representation is an integer multiple
of $|\Gamma|$.  To show this note that the number of times $n_1$ the
trivial representation appears in $\gamma$ must be an integer, and
this is given by
$$n_1={1\over |\Gamma|}{\rm Tr}\left(\sum_{a} \gamma_a
 \right)={1\over
|\Gamma|}{\rm Tr}(\gamma_1)~.$$
Denoting $n_1$ by $n$, we thus
conclude that ${\rm Tr}(\gamma_1)=n |\Gamma|$. Now recall from
representation theory
of groups that the characters ({\it i.e.}, traces)
of elements in a representation uniquely fix the representation
\ref\group{M. Tinkham, ``Group Theory and Quantum Mechanics'' (New York,
McGraw-Hill, 1964)}. We thus conclude that if the condition \pole\
is satisfied we have $n$ copies of the regular representation of $\Gamma$.

The regular representation decomposes into a direct sum of all irreducible
representations ${\bf r}_i$ of $\Gamma$ with degeneracy factors
$n_i=|{\bf r}_i|$. The gauge group is ($N_i\equiv nn_i$)
$$ G=\otimes_i U(N_i)~.$$
The matter consists of Weyl fermions (and scalars) transforming in
bi-fundamentals
$({\bf N}_i,{\overline {\bf N}}_j)$ according to the decomposition
of the tensor product
of ${\bf 4}$ (${\bf 6}$) of $Spin(6)$ with the corresponding representation
(see \lnv\ for details).

In \lnv\ it was shown (using Feynman diagram techniques)
that the one-loop $\beta$-functions vanish for these gauge
theories. Moreover, in ${\cal N}=1$ cases it was shown that the one-loop
anomalous scaling dimensions for matter fields also vanish. This implies that
the two-loop $\beta$-functions also vanish in the ${\cal N}=1$ theories.

In section 3 we will show that in the large $N$ limit all of these theories are
finite to all
loops. The proof there will crucially depend on the fact that the
representation for
the Chan-Paton matrices $\gamma_a$ satisfies the condition \pole.

\newsec{Large $N$ limit and finiteness}

In this section we consider perturbative expansion of the theories constructed
in section 2 which satisfy the condition \trace. We will work in the full
string theory
framework which as we will see is much simpler than the Feynman diagram
techniques.
At the end we will take $\alpha'\rightarrow 0$ limit which amounts to reducing
the theory to the gauge theory subsector.  In this way we will be able to show
directly that at large $N$ these theories
are finite (including the theories without supersymmetry).   This is in accord
with the arguments in \ks\
corresponding to the region where $\lambda$ is large.  However
\ks\ uses the non-trivial conjecture in \malda\ whereas our
arguments are more elementary and apply to the region
where $\lambda$ is small.
Moreover, we will show that in this limit computation of any correlation
function
in these theories reduces to the corresponding computation in the ${\cal N}=4$
Yang-Mills
theory. We will then discuss the subleading ${1\over N}$ corrections.
We will heuristically argue that
at least for the ${\cal N}=1$ case, the theories under
consideration may remain superconformal even for finite $N$. The story with
${\cal N}=0$ theories appears to be more involved due to complications
associated with
closed string tachyons.

\subsec{Conditions for finiteness}

As we discussed in section 2, the gauge group in the theories we are
considering
is $G=\otimes_i U(N_i) (\subset  U(N))$.
To consider finite theories we delete the $U(1)$ factors
(for which there are matter fields charged under them) and consider
$G=\otimes_i SU(N_i)$.
Our goal in this section is to show that in the large $N$
limit this non-Abelian gauge theory is conformal
(including the $U(1)$'s would not have affected this analysis
as it is subleading in $N$).

To establish that a non-Abelian gauge theory is conformal we need to check
three
points: ({\it i}) gauge coupling non-renormalization which amounts to computing
two-point correlators involving gauge bosons; ({\it ii}) vanishing of anomalous
scaling
dimensions (wave-function non-renormalization) which amounts to computing
two-point
correlators involving matter fields; ({\it iii}) non-renormalization of Yukawa
(three-point)
and quartic scalar (four-point) couplings. As far
as perturbation theory is concerned the third point needs to be checked only in
non-supersymmetric theories for in ${\cal N}=1$ theories we have perturbative
non-renormalization theorem for the superpotential.

There are two classes of diagrams we need to consider: ({\it i}) diagrams
without handles;
({\it ii}) diagrams with handles. The latter correspond to closed string loops
and are
subleading in the large $N$ limit. We will therefore discuss these
contributions only
when we turn to subleading ${1\over N}$ corrections.

The diagrams without handles can be divided into two classes: ({\it i}) planar
diagrams
where all the external lines are attached to the same boundary; ({\it ii})
non-planar diagrams
where the external lines are attached to at least two different boundaries. The
latter are subleading in the large $N$ limit.   We will in fact show
the stronger statement that
all the planar diagrams in the orbifolded theory give correlations
which are {\it identical} to that of the parent ${\cal N}=4$ theory,
by showing that all the diagrams containing twisted boundary
conditions identically vanish in this limit.

\subsec{Planar diagrams}

In the case of planar diagrams we have $b$ boundaries with $M$ external lines
attached
to one of them while others are free.
As noted before to leading order in large $N$ we will always consider
the external lines attached to one boundary as depicted in Fig.1.
The amplitude consists of summing over all possible
twisted boundary conditions (homomorphism
of the fundamental group of the planar diagram to $\Gamma$).
This is simply how string theory ensures that the states contributing
to the amplitudes are
properly projected by the action of the orbifold group. The trivial
boundary conditions correspond to the same amplitudes as in the
${\cal N} =4$ case (modulo factors of $1/\sqrt{|\Gamma|}$ which can
be reabsorbed by a redefinition of $\lambda$).  We will now
show that for the diagrams under consideration for all other
boundary conditions the amplitudes are identically zero!
It turns out that all we need to do is to carefully
consider what Chan-Paton factors we get for each string diagram.

\bigskip
%\midinsert
\epsfxsize 2.truein
\epsfysize 2.truein
\centerline{\epsfxsize 4.truein \epsfysize 4.truein\epsfbox{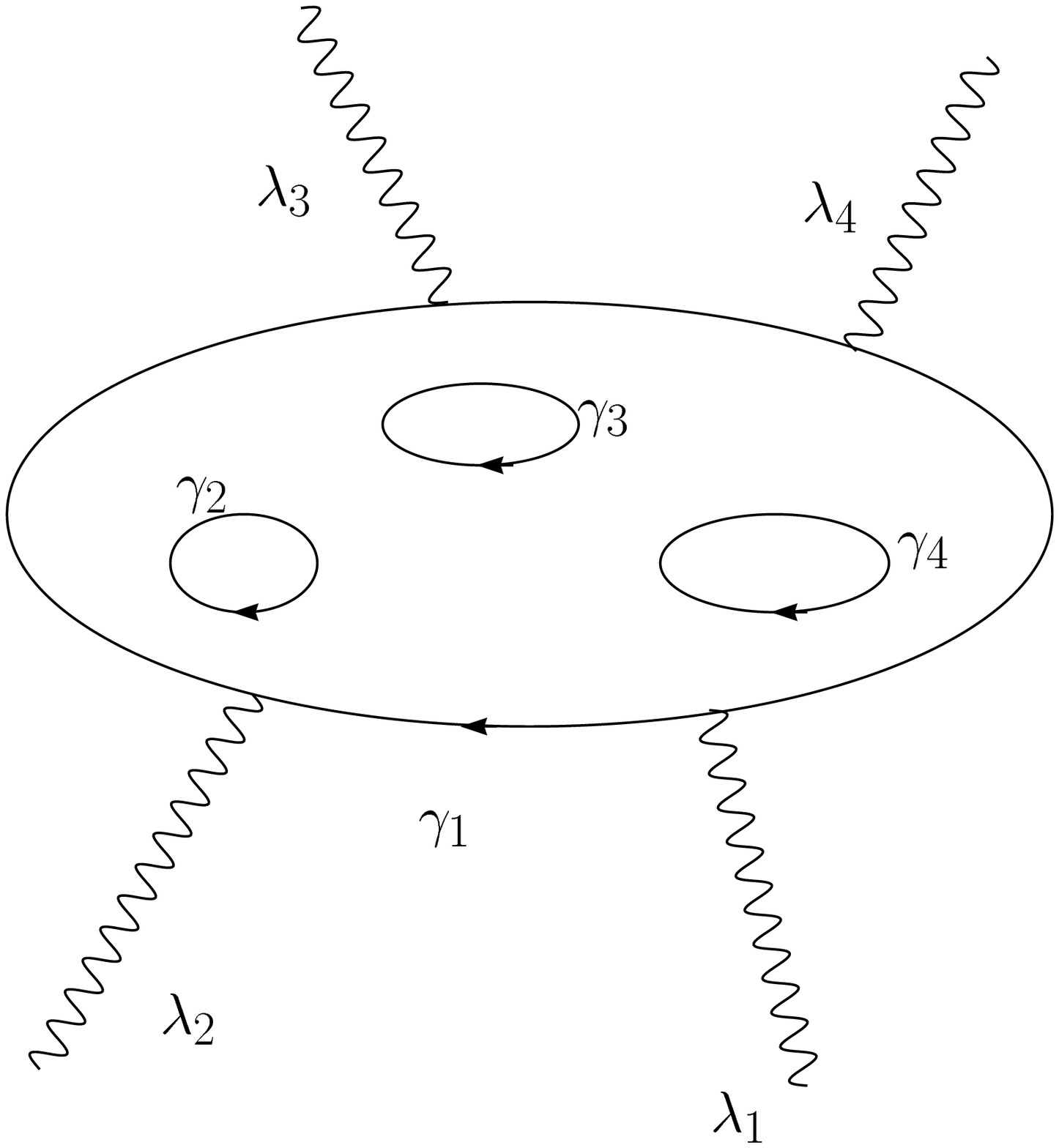}}
\leftskip 2pc
\rightskip 2pc
\vglue.5in
\centerline{\noindent{\ninepoint\sl \baselineskip=8pt {\rm {Fig.1. A planar
diagram.}}}}

Here we need to specify the twists on the
boundaries. A convenient choice (consistent with that made for the annulus
amplitude \partition) is\foot{Here some care is needed in the cases where
$\Gamma$ is non-Abelian and we will have to choose
base points on the world-sheet to define the twists.  However the argument
we give is unmodified also in this case.}
\eqn\mono{\gamma_{a_1}=\prod_{s=2}^{b} \gamma_{a_s}~,
}
where $b$ is the total number of boundaries, $\gamma_{a_1}$ corresponds to the
outer boundary, and $\gamma_{a_s}$, $s=2,\dots b$, correspond to inner
boundaries.

Let $\lambda_r$, $r=1\dots M$, be the Chan-Paton matrices corresponding to the
external lines. Then the planar diagram with $b$ boundaries has the following
Chan-Paton group-theoretic dependence:
$$\sum {\rm Tr}\left(\gamma_{a_1} \lambda_1\dots\lambda_M\right)
\prod_{s=2}^{b}  {\rm Tr}(\gamma_{a_s})~,$$
where the sum involves all possible distributions of $\gamma_{a_s}$ twists
that satisfy the condition \mono\ as well as permutations of $\lambda_r$
factors (note that the $\lambda$'s here are the states
which are kept after projection and so they commute with the action
of $\gamma$'s). The important point here
is that unless all twists $\gamma_{a_s}$ are trivial for $s=2,\dots,b$, the
above
diagram vanishes by the virtue of
\pole. But then from \mono\ it follows that $\gamma_{a_1}$
is the identity element as well.
This is exactly what we wished to prove.  We have thus established
that to leading order in $N$ all the amplitudes of the
orbifolded theories agree with the corresponding ones for
${\cal N}=4$ case (with a simple rescaling
of coupling), and in particular the $\beta$-functions and anomalous
scaling dimensions all vanish.

\subsec{Non-planar diagrams without handles}
We will now consider the extension of the previous considerations
to non-planar diagrams at $g=0$, {\it i.e.}, diagrams obtained by attaching
vertex operators to different boundaries.

Let us start with 2-point functions. Consider a non-planar diagram with
an arbitrary number of boundaries
and two external lines attached to two distinct boundaries.
In order for the amplitude not to be zero we need the twist
along the other boundaries to be trivial. So the only possibility
is that if the two boundaries with external lines have the same twists
(with opposite orientations for the boundaries).
 Thus
the Chan-Paton group-theoretic dependence of this diagram is given by
$$\sum_a {\rm Tr}\left(\gamma_{a} \lambda_1\right)
{\rm Tr}\left(\gamma_{a} \lambda_2\right)~.$$
If $\lambda_1$ or $\lambda_2$ correspond to charged
fields (such as {\it non-Abelian}
gauge bosons), then this expression vanishes. Indeed, for the adjoint of
$SU(N_i)$ we have ${\rm Tr}(\lambda_{1,2})=0$.  Note that this is not
the case
for the neutral fields, such as $U(1)$ gauge bosons
and neutral matter fields (if any).  This is indeed the source
of the running of the coupling constants for $U(1)$'s
(and potential source for anomalous scaling dimensions
for neutral fields).  Thus, we see that non-planar
diagrams without handles do not
contribute to the {\it non-Abelian} gauge coupling renormalization and
anomalous
scaling dimensions of charged matter fields to all loops.
Even though these diagrams are only a subset of subleading corrections
at large $N$, the fact that they vanish for all theories
under consideration (including the non-supersymmetric ones)
is a hint that they may actually be finite even for finite $N$.
This issue will be discussed further below.

We can also extend the above argument to three point functions. In particular
the Yukawa couplings are the same as those of the corresponding
${\cal N}=4$ theory.  But
if we consider higher than three point functions
at genus 0 and put them at various boundaries, there is no
argument why the twisted boundary conditions would not contribute.
This thus shows that not all the correlation functions
are going to be the same as the corresponding ${\cal N}=4$
theory in the subleading corrections.  In particular, the quartic
self-interaction of bosons for the ${\cal N}=0$
case may {\it a priori} be different from that
of the ${\cal N}=4$ case, by attaching
two pairs of bosons to two boundaries, for the non-planar diagrams in
subleading
order in $N$.
The Chan-Paton group-theoretic dependence of this diagram is given by
$$\sum {\rm Tr}\left(\gamma_{a} \lambda_1\lambda_2\right)
{\rm Tr}\left(\gamma_{a} \lambda_3\lambda_4\right)~.$$
Note that this is the case where by
factorization property of string theory it can be represented as two disc
diagrams
connected by a long thin tube corresponding to a closed string exchange.

We can {\it a priori} expect twisted
closed string states, including tachyons, propagating in the tube
to contribute.   The
tachyons, at least naively, may lead to infrared divergences. (In section
4 we discuss some aspects of this.) The twisted massless states may also
contribute to the infrared divergence. Note that the contributions
corresponding
to $a=1$ still look like the ${\cal N}=4$ correlators and are therefore finite.
As for the twisted sector contributions, it is unclear whether they can be
finite
in the $\alpha'\rightarrow 0$ limit.

\newsec{Subleading corrections at large $N$:  diagrams with handles}
So far we have talked about leading order diagrams at large $N$ as
well as some subleading contributions depending on where we insert
the external states to the planar diagram.  We have shown
that the beta functions for all these cases vanish, with
the possible exception of quartic scalar coupling in the ${\cal N}=0$
theories in the subleading order in $N$.  Here we wish to see what
can be said by including other subleading corrections in $N$ which
come from including handles.

Note that now we can have also twists around the handles
which make contributions to amplitudes which
are different from the ${\cal N}=4$ case (which would correspond
to the case without handles).  Thus, {\it a priori} all amplitudes
may have $1/N$ corrections which are distinct from those of the corresponding
${\cal N}=4$ theory.

Let us however ask if the $\beta$-functions will be zero or not, {\it i.e.},
let us see if the theories under consideration will be conformal
at subleading order in $N$, when we include handles.

Here we note that renormalization in effective field theory is due to string
theory
infrared divergences corresponding to massless modes \ref\kap{V. Kaplunovsky,
``One-Loop Threshold Effects in String Unification'', Nucl. Phys. {\bf B307}
(1988) 145;
ERRATA for ``One-Loop Threshold Effects in String Unification'',
Nucl. Phys. {\bf B382} (1992) 436, hep-th/9205068.}. Thus, if a given diagram
does not
contain infrared divergences it will not contribute into the field-theoretic
renormalization of the corresponding correlator. If the diagram is non-zero,
however,
there is a {\it finite} (and {\it independent} of the energy scale)
renormalization. Such renormalizations might change
the value of the fixed point couplings, but not
the fact that we have vanishing beta functions.

The case where $\Gamma \subset SU(2)$, which gives rise to ${\cal N}=2$,
is finite, because there are no higher loop correction to
$\beta$-functions in these theories.  We are thus primarily interested in the
${\cal N}=0,1$ cases.  Let us consider the ${\cal N}=1$ case first.
We have already argued that all the $\beta$-functions vanish for $g=0$
diagrams with arbitrary boundaries.  To consider what happens by
including handles we consider a disc with arbitrary number of holes
and with $g$ handles attached to it.  Let us consider two
point functions (which in the ${\cal N}=1$ case is sufficient
for checking perturbative finiteness).  In this case both vertex operators are
attached to the same boundary (otherwise the trace would vanish).
We are looking at potential sources of infrared divergences in such
amplitudes.  Since we have already argued that the disc with arbitrary
boundaries by itself does not have divergent contributions,
any potential divergences may come from the infrared divergences
in the integration over the handle moduli, or from corners of moduli
where handles approach boundaries of the disc.

The integration over the handle moduli will not have any infrared divergences,
as that would correspond to massless tadpoles for closed type IIB string
theory in a background with ${\cal N}=1$ supersymmetry.  Thus the only
potential
source of divergence is when handles approach boundary points.

We will now present evidence why these may also be zero (modulo
subtleties having to do with neutral fields discussed below).  Let us
first consider the case of a disc with a single handle attached
(Fig. 2)
and consider contributions to the $\beta$-function for gauge theory
(say, from two point function of gauge fields).

\bigskip
%\midinsert
\epsfxsize 2.truein
\epsfysize 2.truein
\centerline{\epsfxsize 4.truein \epsfysize 4.truein\epsfbox{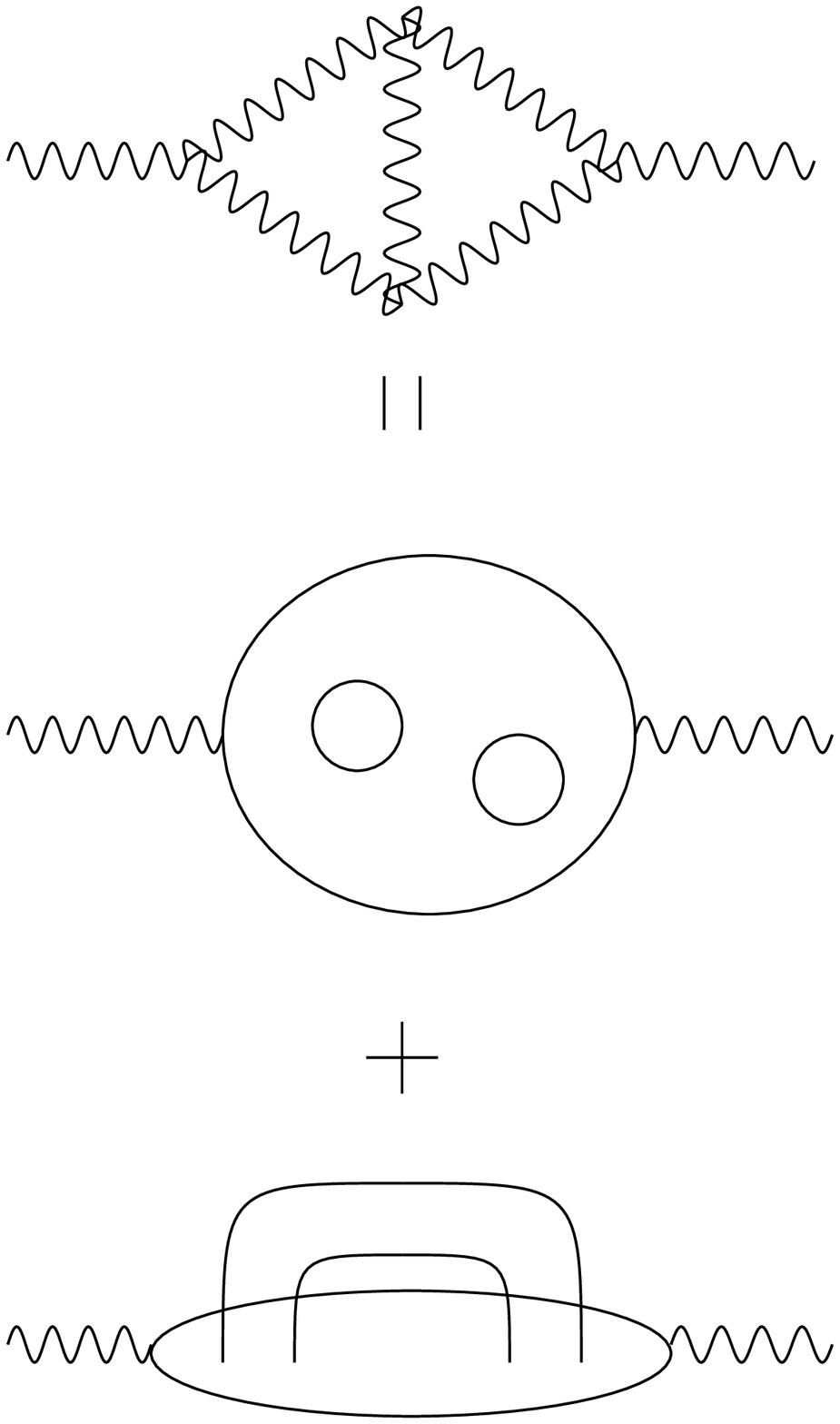}}
\leftskip 2pc
\rightskip 2pc
\vglue .5in
\centerline{\noindent{\ninepoint\sl \baselineskip=8pt {\rm {Fig.2.
Two-loop order in large $N$ limit.}}}}

We consider two
points on the boundary of the disc, corresponding to insertion
of vertex operators of the gauge fields.  In this case the only
potential source of divergence arises when the handle approaches the
boundary (at the insertion points or otherwise).  However we will
now show there is no such divergence.  It was shown
in \lnv\ that in the ${\cal N}=1$ case the $\beta$-functions vanish
to two loops for any $N$.  When translating this to large $N$ order
we have two contributions, one from a $g=0$ diagram with three boundaries,
and the other with a disc and one handle attached.  Thus the sum
of these diagrams will give zero contribution to $\beta$-functions.
On the other hand, we have already shown that the one corresponding to the
$g=0$ diagram does not contribute to the $\beta$-functions.  This shows,
therefore, that the disc with one handle attached will also not
contribute to the $\beta$-functions.

If we have more handles, from what we have said, it is essentially clear that
each individual handle approaching boundaries will not give any
contributions to the $\beta$-function.  The question of what happens
when some number of them approach each other and the boundary is not
completely clear.  However it is likely that even in this case, using
factorization properties of string amplitudes and the structure
of boundary of moduli spaces one may be able to prove that they are zero.
However, we have to note\foot{We would like
to thank Andrei Johansen for pointing this out to us.}\ that in these
computations we are still {\it including} all the neutral fields, and in
particular the $U(1)$'s.  However as already noted the $U(1)$ couplings do run
and the scaling dimensions of neutral fields are non-zero even in the limit
where we ignore handles (but subleading in $N$).  What this means is that if we
drop all the neutral
fields from the spectrum of field theory, in order to get a finite theory,
we will get some shifts of the order of
$1/N^2$ in the values of the coupling constants at the
fixed points inherited from the parent $N=4$ theory.

For the ${\cal N}=0$ case the argument is less clear.  Here we do
have infrared divergences from the closed string sector by itself.  In
particular there is no reason why one point amplitude of the dilaton
tadpole is zero, and that itself will be a potential source of contributions
to $\beta$-functions.  The tachyon is another potential source for infinities.
Nonetheless, one may formally expect them to be irrelevant for the gauge theory
discussion, because we are taking the limit $\alpha'\rightarrow 0$
and so the square of tachyon mass is $-\infty$.  One
would expect to formally subtract them off in gauge theory discussions,
along with massive modes of strings (perhaps a way to make this
precise is to consider $\alpha'\rightarrow i\epsilon$).  Clearly
more work remains to be done in this direction to settle the exact
finiteness of the ${\cal N}=0$ theories.

This works is supported by NSF grant PHY-92-18167.  The research
of M.B. is supported in addition by NSF 1994 NYI award and the DOE
1994 OJI award. The work of Z.K. was supported in part by the grant
NSF PHY-96-02074, and the DOE 1994 OJI award. Z.K. would also like to
thank Albert and Ribena Yu for financial support.

\listrefs

\end